\documentclass[%
 reprint,
 amsmath,amssymb,
 prd, 
]{revtex4-2}
\usepackage{graphicx}
\usepackage{booktabs}
\usepackage{dcolumn}
\usepackage{bm}
\usepackage{amsmath}
\usepackage{geometry} 
\usepackage[hidelinks]{hyperref}
\usepackage{cleveref}
\usepackage{natbib}
\usepackage{aas_macros}
\begin{document}

\preprint{APS/123-QED}

\title{CMB Constraints on Quantized Spatial Curvature $\Omega_K$ in globally CPT-symmetric universes}

\author{Wei-Ning Deng}
\email{wnd22@cam.ac.uk}

\author{Will Handley}%
\email{wh260@cam.ac.uk}
\affiliation{Astrophysics Group, Cavendish Laboratory, J.J. Thomson Avenue, Cambridge, CB3 0HE, UK}
\affiliation{Kavli Institute for Cosmology, Madingley Road, Cambridge, CB3 0HA, UK} 


\date{\today}

\begin{abstract}
The periodic solution of the Friedmann equation in conformal time, implies that only cosmological perturbations exhibiting corresponding symmetries are physically permissible, leading to a discrete spectrum of allowed wave vectors. 
Furthermore, in a spatially closed universe, these wave vectors are independently constrained to be integers. 
Matching these two distinct quantization conditions provides a novel theoretical constraint on the possible values of spatial curvature. 
In this work, we numerically solve the cosmological perturbation equations, incorporating radiation anisotropy and higher-order Boltzmann terms, to calculate these discrete wave vectors with improved precision. 
Subsequently, we generate Cosmic Microwave Background (CMB) power spectra for different characteristic spacings of these quantized wave vectors. 
Finally, we apply the constraint to \texttt{Planck 2018} observational data to determine the cosmological parameters. This analysis yields a discrete set of allowed values for the spatial curvature, $\Omega_K$, including $[-0.076,-0.039, -0.024, -0.016, -0.012, \dots]$. 
\end{abstract}

\maketitle

\section{Introduction}

The spatial curvature of our universe, denoted by $\Omega_K$, has been a subject of wide interest and ongoing investigation. Data from \texttt{Planck 2018} indicate a preference for a closed universe, with $-0.095 < \Omega_K < -0.007$ at $99\%$ confidence level  \cite{2020A&A...641A...6P}. However, the inclusion of other observational datasets, such as those from Baryon Acoustic Oscillations (BAO) and \texttt{Planck}'s lensing, tend to favor a spatially flat universe \cite{2020A&A...641A...6P, PhysRevD.105.063524}.

Nevertheless, this conclusion has been challenged in recent literature. Studies by Di Valentino et al. \cite{2020NatAs...4..196D} and Handley \cite{2021PhRvD.103d1301H} demonstrate Bayesian evidence supporting a closed universe. This tension is often attributed to a disagreement between the lensing amplitude ($A_L$) and the implicit assumption of a fiducial $\Lambda$CDM cosmology within BAO analyses. The preference for a flat universe can be significantly weakened when these underlying assumptions are relaxed \cite{2022MNRAS.517.3087G}.

If the universe is indeed closed, its finite spatial volume imposes a quantization condition, restricting the allowed comoving wave vectors to integer multiples of a fundamental mode related to the curvature radius. 
However, this is not the sole potential constraint on wave vectors. Solving the Friedmann equation in conformal time, rather than physical time, reveals a periodic solution for the scale factor \cite{2021arXiv210906204B, 2024PhLB..84938442B}.
Considering this periodicity show important insight to our universe. For example, it led Boyle and Turok to further postulate that the universe is charge, parity, and time (CPT) symmetric at the Big Bang \cite{2021arXiv211006258B,2022arXiv220810396B,2024PhLB..84938442B,2023arXiv230200344T}. Such a CPT-symmetric framework offers novel explanations for several outstanding problems in cosmology and particle physics without necessarily introducing additional model complexity. 
On the other hand, Lasenby also investigated the evolution of cosmological perturbations within such a periodic framework and found that physical consistency requires these perturbations to also be periodic, thereby constraining their wave vectors to a discrete set of values \cite{2022PhRvD.105h3514L}.

In \cite{PhysRevD.110.103528}, we pointed out that the constraints on wave vectors arising from a spatially closed geometry and those from the periodicity of the universe must be consistent. This consistency requirement imposes a strong theoretical constraint on the value of spatial curvature, implying that only specific, discrete values of $\Omega_K$ are permissible. Building upon this foundation, the present work incorporates more realistic cosmological components, namely matter ($\Omega_m$) and radiation anisotropy. To achieve this, we adapt the numerical solutions in \cite{2022PhRvD.105l3508P}, and modify it to include spatial curvature. By subsequently imposing the quantization condition from our previous work, we determine the permissible curvature values within this more realistic model. The final step is to compare these discrete values with observational data to identify which are empirically viable.

We begin by reviewing the underlying theory in \cref{sec:theory}. We then numerically solve for the discrete wave vectors within a realistic cosmological model in \cref{sec:real_cosmology}. Subsequently, in \cref{sec:CMB}, we generate the corresponding CMB power spectra and compare them with \texttt{Planck 2018} data. Our final results and discussion are presented in \cref{sec:result}. 

\section{Theory}\label{sec:theory}

Let us first review the formulation of the theory. Throughout this paper, we adopt natural units where $8\pi G=c=\hbar=k_B=1$. 

\subsection{Perturbation equations and solution} 

The perturbed FLRW metric in the Newtonian gauge can be written as: 
\begin{multline}
    \label{eq:metric}
    ds^2 = a^2(\eta)\bigl[-(1 + 2\Psi)d\eta^2 \\
    + (1 - 2\Phi)\left(\frac{dr^2}{1-\kappa r^2}+r^2d\Omega^2\right)\bigr]
\end{multline}
where $a(\eta)$ is the scale factor as a function of conformal time $\eta$, $\Phi$ and $\Psi$ are the scalar metric perturbations, and $\kappa$ is the spatial curvature parameter.  

The zeroth-order Friedmann equations in conformal time can be expressed in dimensionless form as: 
\begin{equation}\label{eq:friedmann_eq}
\begin{aligned}
    3\dot{a}^2 = \lambda (a^4 - 3\tilde\kappa a^2  + \tilde ma+\tilde r), \\
    6\ddot{a} = \lambda(4a^3-6\tilde\kappa a+\tilde m),
\end{aligned}
\end{equation}
where $\dot{a} \equiv da/d\eta$, and $\tilde{\kappa} = \kappa/\lambda$, $\tilde{m} = m/\lambda$, $\tilde{r} = r/\lambda$ are dimensionless parameters representing curvature, matter, and radiation energy densities, respectively, scaled by a parameter $\lambda$ related to the cosmological constant. 

Lasenby solved the first-order perturbation equations within the periodic framework and found that, for physical consistency (avoiding divergences or unphysical behavior in the "anti-universe" phase), the perturbation solutions must be periodic \cite{2022PhRvD.105h3514L}. 
These solutions can be either symmetric or anti-symmetric at $\eta=\eta_\infty$ (the "end" of the universe in one cycle), leading to the quantization condition $\sqrt{3}\tilde k\eta_\infty=n\pi/2$, for $n \in \mathbb{N}$, where $\tilde k = k_{\mathrm{com}}/\sqrt{\lambda}$ is the dimensionless comoving wave number. 
However, Lasenby also noted a complication: the matter density, $\rho_m \propto 1/a^{3}$, would lead to negative value in the anti-universe phase (where $a<0$ in the periodic mathematical extension). 
Such negative energy density is problematic and would also break the symmetry of the background evolution. 
To address this, Boyle proposed that the matter density should scale as $\rho_m \propto 1/|a|^3$. This modification arises because the classical fluid approximation $\rho_m \propto a^{-3}$ is expected to break down near the Big Bang singularity and the end of the universe, where a quantum field theoretical description becomes necessary. For a scalar field, it is easy to check that the solution (of Klein-Gordon) maintains symmetry in both situations. 
However, for fermionic matter, which constitutes standard matter, is more complex due to spin connection terms in the Dirac equation in curved spacetime; a full treatment is deferred to future work. 

If one accepts the $\rho_m \propto 1/|a|^3$ scaling, the evolution of the universe and its anti-universe becomes symmetric. In this scenario, only perturbation modes that are symmetric with respect to the Bang and $\eta_\infty$ are considered physically admissible. 
This restricts the quantization condition to $\sqrt{3}\tilde k\eta_\infty=n\pi$, for $n \in \mathbb{N}$ (\cite{2022PhRvD.105h3515B}). 
The discrete dimensionless wave vectors then have a characteristic spacing. 

\subsection{Basic idea}

In our previous work \cite{PhysRevD.110.103528}, we extended Lasenby's analysis \cite{2022PhRvD.105h3514L} to spatially curved universes. 
Spatial curvature affects the conformal age of the universe, $\eta_\infty$, and consequently, the spacing of the discrete dimensionless wave vectors, $\Delta \tilde k$, becomes dependent on this curvature. (Recall that $\tilde k \equiv k_{\mathrm{com}}/\sqrt{\lambda}$.) 

For a spatially closed universe ($\kappa > 0$), the finite spatial volume independently quantizes the allowed comoving wave numbers. Specifically, $k \equiv k_{\mathrm{com}}/\sqrt{\kappa}$ must effectively be an integer, $k \in \mathbb{N}$. 
This implies that the dimensionless wave vectors $\tilde k = k_{\mathrm{com}}/\sqrt{\lambda}$ are quantized as $\tilde k_j = j \sqrt{\tilde{\kappa}}$ for some integer $j$, leading to a characteristic spacing related to $\sqrt{\tilde{\kappa}}$. 
For consistency between the periodic-derived quantization and the geometric quantization, the periodic-derived spacing $\Delta \tilde k = N \sqrt{\tilde{\kappa}}$, where $N \in \mathbb{N}$. 
Requiring such consistency restricts the possible values of spatial curvature, $\tilde{\kappa}$ (and thus $\Omega_K$), to a discrete set, as demonstrated in \cite{PhysRevD.110.103528}.
In this work, we extend this framework to a realistic cosmological model and aim to identify the observationally favoured discrete curvature values by comparing these theoretical predictions with CMB data. 

\section{Realistic Cosmological Model}\label{sec:real_cosmology} 

To apply this model to the observable universe, we first establish the transformation between the dimensionless parameters of the theory ($\tilde m, \tilde \kappa, \tilde r$) and standard cosmological density parameters ($\Omega_m, \Omega_K, \Omega_r, \Omega_\Lambda$). 
Subsequently, we calculate the periodic-derived wave vector spacing, $\Delta \tilde k$, by incorporating the effects of radiation anisotropy and higher-order terms in the Boltzmann hierarchy, adapting the methods of \cite{2022PhRvD.105h3515B, 2022PhRvD.105l3508P}. 

\subsection{Parameter transformation}\label{subsec:param_transform}

We begin by relating the dimensionless parameters $\tilde r, \tilde m, \tilde \kappa$ to the present-day cosmological density parameters $\Omega_\Lambda, \Omega_r, \Omega_m, \Omega_K$. 
The transformations are given by:
\begin{equation}\label{eq:transformation}
    \Omega_m=\tilde m\Omega_\Lambda/a_0^3,\quad \Omega_K=-3\tilde \kappa\Omega_\Lambda/a_0^2,\quad \Omega_r=\tilde r\Omega_\Lambda/a_0^4,
\end{equation}
where $a_0$ is the present-day scale factor.
Using the Friedmann constraint equation at the present epoch, $\Omega_\Lambda+\Omega_K+\Omega_m+\Omega_r=1$, and substituting the definitions from Eq.~\eqref{eq:transformation}, we obtain (after dividing by $\Omega_\Lambda$): 
\begin{equation}\label{eq:friedmann_constraint_dimless}
    1-3\frac{\tilde\kappa}{a_0^2}+\frac{\tilde m}{a_0^3}+\frac{\tilde r}{a_0^4}=\frac{1}{\Omega_\Lambda}.
\end{equation}
Without loss of generality, we set $\tilde r=1$. This choice normalizes the scale factor such that $a=1$ at the epoch where the energy density contribution from $\lambda$ equals that from radiation (i.e., $r/a^4 = \lambda$, so $a^4 = r/\lambda = \tilde r$). 
Consequently, from the definition of $\Omega_r$, the present-day scale factor $a_0$ is given by $a_0=(\Omega_\Lambda/\Omega_r)^{1/4}$ (since $\Omega_r = (1 \cdot \Omega_\Lambda)/a_0^4$ with $\tilde r=1$).
Equation \eqref{eq:friedmann_constraint_dimless} then becomes a quartic equation for $a_0$: 
\[
a_0^4 -3\tilde\kappa a_0^2+\tilde ma_0 +\left(1-\frac{1}{\Omega_r}\right)=0. 
\]
Solving this quartic equation for $a_0$ yields four roots; the physically relevant solution is the unique positive real root.
We choose to fix the present-day radiation density parameter, $\Omega_r$, as its value is relatively well-determined. We assume a photon density parameter $\Omega_\gamma h^2 = 2.47\times 10^{-5}$ and an effective number of relativistic species $N_{\mathrm\mathrm{eff}}=3.046$, so that: 
\begin{equation}
    \Omega_r= (1 + N_{\mathrm{eff}}(7/8)(4/11)^{4/3} ) \Omega_\gamma.
\end{equation}
With $\Omega_r$ fixed and $a_0$ determined, we calculate $\Omega_\Lambda=a_0^4\Omega_r$. Subsequently, $\Omega_K$ and $\Omega_m$ are found using Eq.~\eqref{eq:transformation}. 
In summary, given values for $\tilde \kappa$, $\tilde m$ (with $\tilde r=1$) and the observational value of $\Omega_r$ (which depends on $h$), we can determine $a_0$ and subsequently the full set of standard density parameters $\Omega_\Lambda, \Omega_K, \Omega_m$.

\subsection{Calculating $\Delta \tilde k$ considering radiation anisotropy and higher-order terms} 

To accurately calculate the periodic-derived wave vector spacing $\Delta \tilde k$, we must solve the cosmological perturbation equations, identify the symmetric solutions, and determine the spacing between their allowed wave vectors. 
Following the methodology of \cite{2022PhRvD.105h3515B}, we evolve the perturbations using a perfect fluid approximation up to the epoch of recombination. After recombination, we incorporate radiation anisotropy by solving the Boltzmann hierarchy for photons. This involves tracking the momentum-averaged Legendre moments of the photon distribution function, $F_{r\ell}$, and its polarization components, $G_{r\ell}$. 

We begin by modifying the perturbation equations (equations~(7)-(15) in \cite{2022PhRvD.105h3515B}) to include terms for non-zero spatial curvature, based on the formalism presented in, e.g., \cite{1992PhR...215..203M}:
\begin{equation}\label{eq:coupled_eqns1}
\begin{aligned}
    &(k_{\mathrm{com}}^2-3\kappa)\Phi =-\frac{1}{2}a^2\sum_i (\delta\rho_i-3\frac{\dot a}{a}(\bar{\rho}_i+\bar{P}_i)v_i), \\ 
    \dot\Phi &=-\frac{\dot a}{a}\Psi-\frac{1}{2}a^2\sum_i(1+w_i)\rho_i v_i, \\ 
    &k_{\mathrm{com}}^2(\Phi-\Psi) =\frac{3}{2}a^2\sum_i (\bar{\rho}_i+\bar{P}_i)\sigma_i, \\ 
    \dot\delta_i &=(1+w_i)( 3\dot\Phi+v_i k_{\mathrm{com}}^2) , \\ 
\end{aligned}
\end{equation}
\begin{equation}\label{eq:coupled_eq2}
\begin{aligned} 
\dot v_i = 3\frac{\dot a}{a}(w_i-\frac{1}{3})v_i-\frac{w_i}{1+w_i}(\delta_i-\frac{2}{3}(1-\frac{3\kappa}{k_{\mathrm{com}}^2})\Pi_i)\\-\Psi-\frac{an_e\sigma_T}{k_{\mathrm{com}}^2}(\theta_b-\theta_r). 
\end{aligned}
\end{equation}
Here, $\theta_i = -k_{\mathrm{com}}^2 v_i$ (for scalar velocity $v_i$), $\sigma_i$ is the anisotropic stress, $\Pi_i$ is related to $\sigma_i$ (e.g., $\sigma_i = \frac{1}{6}\Pi_i$ or similar for curved space), $n_e$ is the proper mean density of free electrons, and $\sigma_T$ is the Thomson scattering cross section. 

After recombination, we assume photons are free-streaming. In this regime, the Thomson scattering term proportional to $n_e\sigma_T$ vanishes. 
With this simplification, the photon polarization moments $G_{r\ell}$ decouple from the intensity moments $F_{r\ell}$ and other perturbations, and are thus neglected for calculating $\Delta \tilde k$. 
The evolution equations for $F_{r\ell}$ are: 
\begin{equation}\label{eq:coupled_eq3}
    \dot{F}_{r\ell} =\frac{k_{\mathrm{com}}}{2\ell+1}[\ell F_{r(\ell -1)}-(\ell +1)F_{r(\ell+1)}] - n_e\sigma_T |a| F_{r\ell}.
\end{equation}
Note the relations $F_{r0}=\delta_r$ (density perturbation), $F_{r1} \propto k_{\mathrm{com}}v_r$ (dipole/velocity), and $F_{r2} \propto \Pi_r$ (quadrupole/anisotropic stress), which show how the macroscopic radiation fluid perturbations are coupled to the moments $F_{r\ell}$. 

We then follow the procedure in \cite{2022PhRvD.105l3508P} to calculate the linear evolution mapping for perturbations from recombination to the end of the universe ($\eta_\infty$). 
This mapping, combined with the symmetry conditions imposed at $\eta_\infty$ and the perfect fluid solutions at recombination, allows determination of the allowed perturbation modes. Calculating this mapping involves integrating the perturbation equations backward in time from near $\eta_\infty$ to recombination. To mitigate numerical instabilities near $\eta_\infty$, the backward integration is initiated from $\eta'=\eta_\infty-\Delta \eta$. The solutions at $\eta'$ are then extrapolated to $\eta_\infty$ using a Taylor expansion, whose coefficients are derived from the coupled perturbation equations \eqref{eq:coupled_eqns1}-\eqref{eq:coupled_eq3}. Please refer to appendix~\cref{sec:Taylor} for the coefficients. 

It is noteworthy that \cite{2022PhRvD.105l3508P} performed calculations using fixed \texttt{Planck 2018} best-fit cosmological parameters \cite{2020A&A...641A...6P}. In contrast, this analysis treats these parameters as free, enabling direct comparison with observational data. 
In dimensionless models, $\Delta\tilde k$ depends only on $\tilde m$ and $\tilde \kappa$. Standard cosmological parameters become relevant when converting $\tilde k$ to physical wave vectors $\Delta k_{\mathrm{phys}}= \Delta\tilde k\sqrt{\lambda}/a_0$, where
\[ \lambda = \frac{3H_0^2\Omega_\Lambda}{8\pi G 
}, \qquad a_0=(\Omega_\Lambda/\Omega_r)^{1/4}. \]
However, in our more realistic treatment (perfect fluid pre-recombination, Boltzmann hierarchy post-recombination), $\Delta\tilde k$ itself acquires a dependence on parameters affecting recombination, such as recombination redshift $z_{\mathrm{rec}}$, which is a function of $\Omega_b h^2$ and $H_0 = 100h \text{ km s}^{-1}\text{Mpc}^{-1}$, in addition to $\tilde m$ and $\tilde \kappa$. Note that while $z_{\mathrm{rec}}$ is also dependent on $\Omega_rh^2$, this parameter is held fixed in this work (see \ref{subsec:param_transform}). 
Consequently, our modified computational framework yields $\Delta\tilde k(\tilde m, \tilde \kappa, \Omega_b, h)$. 

\section{CMB Power Spectrum and Comparison with Planck Data}\label{sec:CMB}

Having established the theoretical framework and the method for calculating the discrete wave vector spacing $\Delta\tilde{k}$, we now turn to comparing the model with observational data. Our procedure involves two main steps: first, generating theoretical Cosmic Microwave Background (CMB) power spectra that incorporate the required wave vector discretization, and second, comparing these spectra with \texttt{Planck 2018} data to identify the cosmological parameters that satisfy our model's consistency condition.

\subsection{Modification of the Boltzmann Angular Power Spectrum Code}\label{subsec:modify_CLASS}

To compute the CMB power spectra for our model, we modified the Boltzmann solver \texttt{CLASS} \cite{DiegoBlas_2011}. The standard code integrates over a continuous spectrum of comoving wave numbers, $k_{\mathrm{com}}$. For a closed universe within our theoretical framework, however, only a discrete set of modes is physically permissible. The consistency between the geometric quantization (due to finite volume) and the periodicity-induced quantization (from the CPT-symmetric solution) requires the dimensionless wave number $k \equiv k_{\mathrm{com}}/\sqrt{\kappa}$ to be restricted to integer multiples of a fundamental integer $N$. That is, the allowed modes are given by $k_{j'} = j' N$ for $j' \in \mathbb{N}$. Our modification to \texttt{CLASS} therefore restricts the computation of the CMB power spectrum to this discrete basis of comoving wave numbers, $k_{com, j'} = j' N \sqrt{\kappa}$.

\begin{figure*}
     \includegraphics[width=0.8\textwidth]{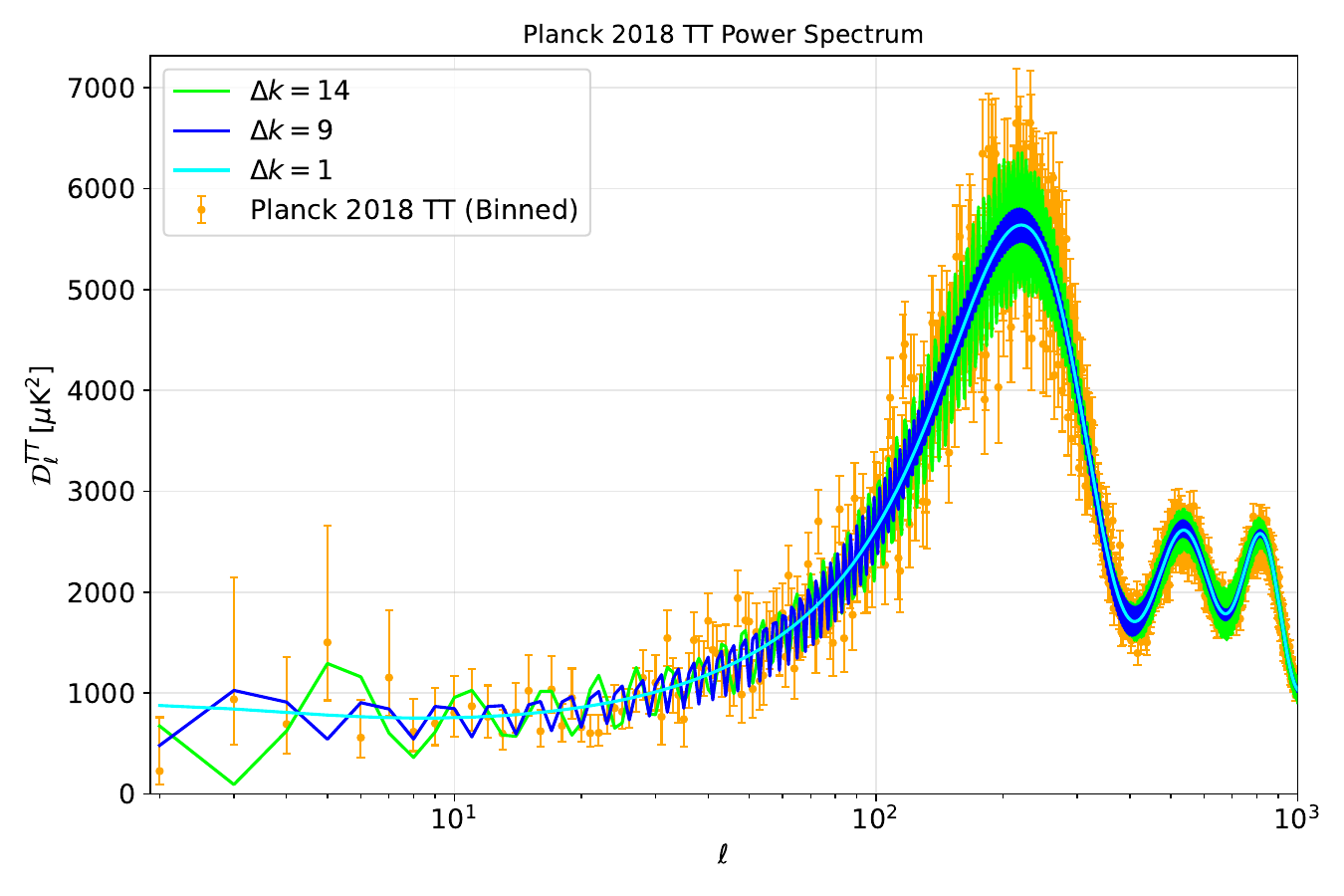}
     \caption{The CMB temperature power spectrum calculated for different effective fundamental mode spacings, characterized by $\Delta k = N$. Here, $N=1$ corresponds to the standard closed $\Lambda$CDM model. For this illustrative plot, we fix $\Omega_K$ and other cosmological parameters to be Planck best-fit values, and isolate the effect of the mode spacing. Larger values of $N$ (a sparser sampling in $k$-space) lead to more fluctuations on large angular scales.}
     \label{fig:CMB_diffN}
\end{figure*}

The value of $\kappa$ itself is not a free parameter for a given $N$, but is determined by solving the consistency condition $\Delta\tilde{k}(\tilde{m}, \tilde{\kappa}, \Omega_b, h) = N\sqrt{\tilde{\kappa}}$. The spacing of the physical modes, and thus the potential for observable discreteness features in the CMB power spectrum, depends on the value of the fundamental mode $N\sqrt{\kappa}$. As illustrated in \cref{fig:CMB_diffN}, a sparser sampling of modes can induce more fluctuating features in the CMB power spectrum.

However, a key aspect of our model mitigates this effect. The value of $\Delta\tilde{k}$ is primarily determined by the conformal age of the universe, $\eta_\infty$, which is not expected to vary dramatically across the observationally viable parameter space. The constraint $\Delta\tilde{k} = N\sqrt{\tilde{\kappa}}$ therefore implies an inverse relationship: larger integer values of $N$ necessitate smaller values of $\tilde{\kappa}$ (and consequently, smaller magnitudes of spatial curvature $|\Omega_K|$). This ensures that the effective fundamental mode in dimensionless units, $N\sqrt{\tilde{\kappa}} = \Delta\tilde{k}$, remains relatively constant. As a result, the physical fundamental mode, $k_{\mathrm{phys}} = (N\sqrt{\kappa})/a_0 = \Delta\tilde{k}\sqrt{\lambda}/a_0$, does not become excessively large, and the fluctuation in the CMB spectrum remain subtle for the parameter space favored by observations.

\subsection{Post-processing \texttt{Planck 2018} MCMC Samples}\label{subsec:compare_planck}

Given that the discreteness features are not expected to be prominent, we can employ an efficient method to constrain our model's parameters without performing a full, computationally expensive Markov Chain Monte Carlo (MCMC) analysis. Instead of fitting our modified \texttt{CLASS} code directly to the data, we post-process the publicly available \texttt{Planck 2018} MCMC chains from the standard $\Lambda$CDM+$\Omega_K$ analysis (\texttt{base\_omegak\_plikHM\_TTTEEE\_lowl\_lowE}). This approach is valid because the likelihood surface of the standard model serves as a good approximation for our model.

Our procedure is as follows: for each sample point in the Planck MCMC chain, we use its cosmological parameters ($\Omega_m, \Omega_K, \Omega_b, h$, etc.) to calculate $\Delta k \equiv \Delta\tilde{k}(\tilde{m}, \tilde{\kappa}, \Omega_b, h) / \sqrt{\tilde{\kappa}}$. We then filter the chains, retaining only those samples for which this calculated value is approximately equal $\pm 2\%$ to an integer $N$. This filtering process isolates the regions of the standard model's posterior distribution that are consistent with our theoretical constraint. These filtered subsamples are then used to construct the marginalized posterior distributions for each allowed integer $N$, which we visualize using \texttt{anesthetic} \cite{anesthetic}.

\section{Results and Discussion}\label{sec:result}

\begin{figure*}
     \centering 
     \includegraphics[width=\textwidth]{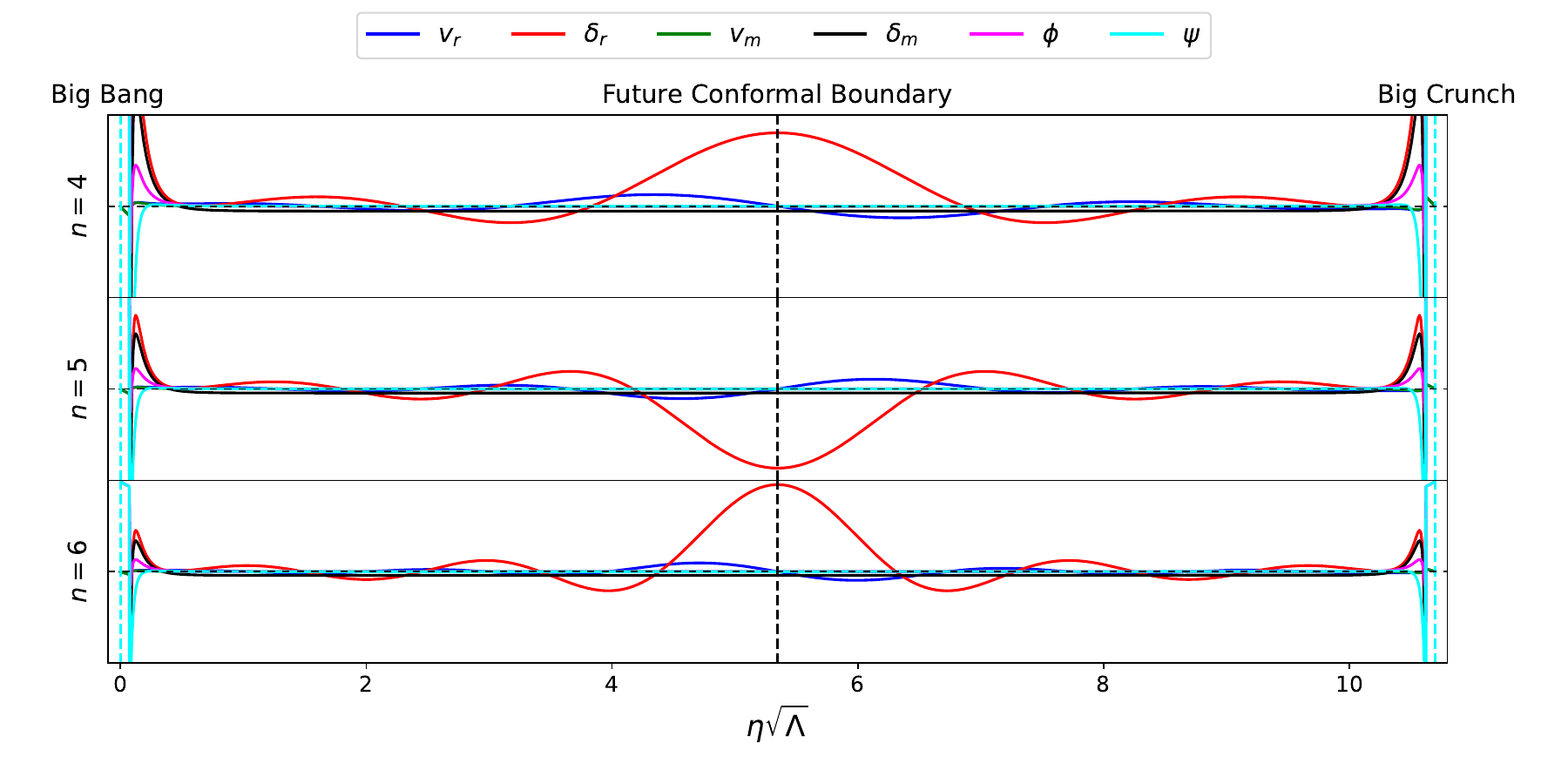}
     \caption{Numerical solutions for several allowed perturbation modes (modes 4--6 are shown to clarify the oscillatory behavior). As expected, the solutions are totally symmetric or antisymmetric at the future conformal boundary. The plot uses best-fit parameters corresponding to the $\Delta k=4$ case: $\Omega_m=0.435, \Omega_K=-0.038, h=0.564, \Omega_b=0.072$.}
     \label{fig:perturbations_allowedK}
\end{figure*}

\begin{figure}
     \includegraphics[width=0.5\textwidth]{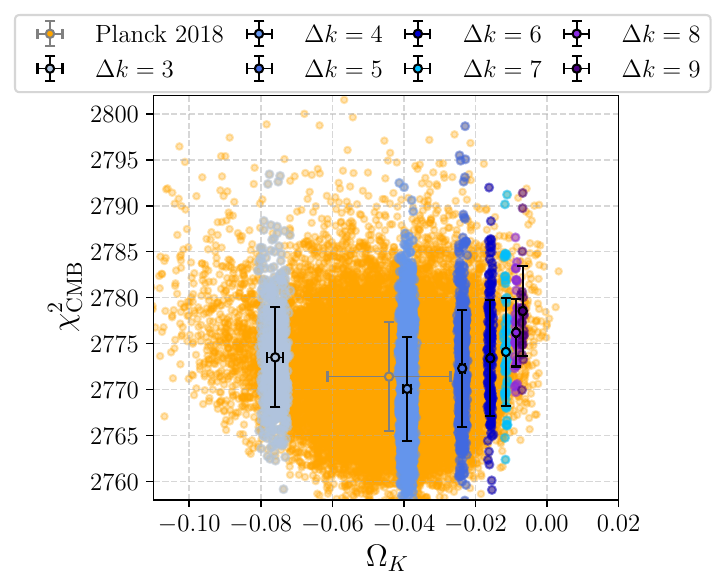}
     \caption{A scatter plot of $\Delta\chi^2$ versus $\Omega_K$ for the filtered MCMC samples. The points show the mean value and $1\sigma$ error bars for each integer solution $\Delta k=N$. The minimum $\Delta\chi^2$ occurs for $N=4$, indicating it is the solution most favored by the \texttt{Planck 2018} data.}
     \label{fig:omegak_vs_chi2}
\end{figure}

\Cref{fig:perturbations_allowedK} displays the numerical solutions for several allowed perturbation modes. As theoretically expected, these solutions exhibit total symmetry or antisymmetry at the future conformal boundary. The solutions are generated using the best-fit cosmological parameters for the $\Delta k=4$ case, a choice we justify later as being favored by \texttt{Planck 2018} data \cref{fig:omegak_vs_chi2}.

\begin{figure*}
     \includegraphics[width=0.85\textwidth]{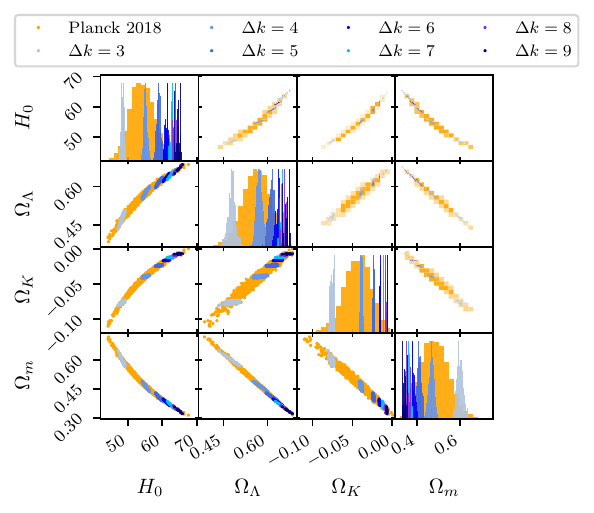}
     \caption{Marginalized posterior distributions from \texttt{Planck 2018} MCMC samples for the \texttt{base\_omegak\_plikHM\_TTTEEE\_lowl\_lowE} likelihood combination. The orange contours represent the full posterior for the standard $\Lambda$CDM+$\Omega_K$ model. The colored regions highlight the subsets of samples that satisfy our consistency constraint $\Delta k \approx N$ for various integers $N$.}
     \label{fig:Planck_diffN}
\end{figure*}

These perturbation solutions are a crucial input for our analysis. Specifically, they are used to compute the function $\Delta\tilde k(\tilde m, \tilde \kappa, \Omega_b, h)$, which in turn allows us to filter the \texttt{Planck 2018} MCMC data samples by selecting only those that satisfy the theoretical constraint $\Delta \tilde k/\sqrt{\tilde{\kappa}}=N$. The outcome of this filtering process, which constitutes the primary result of our analysis, is shown in \cref{fig:Planck_diffN}. This figure presents the marginalized posterior distributions derived from the constrained MCMC samples. The full posterior for the standard $\Lambda$CDM+$\Omega_K$ model is shown in orange, while the colored contours highlight the subsets of samples that satisfy our consistency condition, $\Delta k \approx N$, for different integer values of $N$. The plot clearly demonstrates that our theoretical framework restricts the allowed parameter space to a series of discrete "islands," most prominently visible in the posterior for the spatial curvature parameter, $\Omega_K$. This corresponds to a discrete set of allowed values for the spatial curvature, with the most favored values being $\Omega_K \in [-0.076,-0.039, -0.024, -0.016, -0.012, \dots]$ for $N=3,4, 5, 6, 7, \dots$, respectively.

As predicted by our theoretical considerations in \cref{subsec:modify_CLASS}, the results show a distinct inverse relationship between the integer $N$ and the magnitude of the spatial curvature $|\Omega_K|$. Larger values of $N$ correspond to solutions that are closer to a spatially flat universe. This is a direct consequence of the constraint $\Delta\tilde{k} = N\sqrt{\tilde{\kappa}}$ coupled with the relative stability of the calculated $\Delta\tilde{k}$ across the parameter space. 

To determine which of these discrete parameters is most favored by the data, we examine the $\chi^2$ value for each family of parameters. \cref{fig:omegak_vs_chi2} plots $\Delta\chi^2$ against $\Omega_K$ for the filtered samples. The analysis reveals a clear minimum at $N=4$, indicating that this is the most probable solution according to the \texttt{Planck 2018} data alone. 

It is important to discuss the cosmological parameters associated with the best-fit solution ($N=4$). The preferred values, such as $\Omega_m =0.467^{+0.012}_{-0.013}$ and $H_0 = 55.045^{+0.518}_{-0.485}$ km s$^{-1}$Mpc$^{-1}$, are different from the values derived from a flat $\Lambda$CDM cosmology and other cosmological probes. This discrepancy is a well-known consequence of the strong degeneracy within the CMB data: a wide range of parameter combinations, particularly involving $\Omega_K$, $\Omega_m$, and $H_0$, can produce nearly identical CMB power spectra.

To break this degeneracy and robustly test our model, it is necessary to perform a joint analysis with other cosmological datasets, such as Baryon Acoustic Oscillations (BAO) and Type Ia Supernovae (SNIa). However, such an analysis requires careful consideration of underlying assumptions. Many BAO analyses, for instance, assume a fiducial flat cosmology, which can inadvertently bias the results against a curved universe. A rigorous test demands that spatial curvature be treated as a free parameter consistently across all datasets. Indeed, analyses that do so, such as the work by Glanville et al. \cite{2022MNRAS.517.3087G} on BAO data, have also shown a preference for a closed universe. A comprehensive joint analysis, which we defer to future work, will therefore be crucial for discriminating between the allowed integer solutions and determining the true quantized value of spatial curvature.

\begin{acknowledgments}
WND was supported by a Cambridge Trusts Taiwanese Scholarship. WJH was supported by a Royal Society University Research Fellowship. This work was performed on the DiRAC@Cambridge service of the STFC DiRAC HPC Facility (www.dirac.ac.uk), operated by the University of Cambridge and funded by STFC grants ST/P002307/1, ST/R002452/1, and ST/R00689X/1. DiRAC is part of the National e-Infrastructure. 
\end{acknowledgments}

\appendix
\section{Taylor Expansion Coefficients}\label{sec:Taylor}
To determine the cosmological perturbation solutions at the end of the universe, denoted by $\mathbf{X}^\infty$, we employ a Taylor expansion. For this purpose, it is advantageous to re-express the governing perturbation equations in terms of the reciprocal scale factor, $s \equiv 1/a$. This is because the end of the universe (where the scale factor $a \to \infty$) corresponds to $s \to 0$, a limit around which the coefficients of the differential equations behave well, allowing for a systematic expansion. The solutions at $s=0$ (conformal time $\eta_\infty$) are then extrapolated from solutions at a slightly earlier conformal time $\eta' = \eta_\infty - \Delta\eta$, where $\Delta\eta$ is a small, positive interval and serves as the expansion parameter. The relationship between the solutions at $\eta'$ and $\mathbf{X}^\infty$ is given by:
\begin{equation}\label{eq:taylor_relation}
\begin{pmatrix}
\mathbf{x}' \\
\mathbf{y}'_{2:4}
\end{pmatrix}
=
\begin{pmatrix}
\mathbf{X}_1 \\
\mathbf{X}_2
\end{pmatrix}
\mathbf{X}^\infty
\end{equation}
Here, $\mathbf{X}^\infty$ represents the vector of perturbation modes at $\eta_\infty$ (note that $\phi^\infty=\psi^\infty=0$)
\begin{equation}\label{eq:X_infty_def}
    \mathbf{X}^\infty = (\delta_r^\infty, \delta_m^\infty, v_r^\infty, \dot{v}_m^\infty)^T.
\end{equation}
The vectors $\mathbf{x}'$ and $\mathbf{y}'_{2:4}$ contain the perturbation variables at $\eta'$:
\begin{equation}\label{eq:xy_prime_def}
\begin{aligned}
    \mathbf{x}' &= (\phi', \psi', \delta_r', \delta_m', v_r', v_m')^T, \\
    \mathbf{y}'_{2:4} &= (F_{r2}', F_{r3}')^T,
\end{aligned}
\end{equation}

The matrices $\mathbf{X}_1$ and $\mathbf{X}_2$ contain the Taylor expansion coefficients.

The coefficients within these matrices depend on various cosmological parameters and the wavenumber $k$. We define the following dimensionless combinations. 
\begin{equation}\label{eq:coeffs_c}
c_1 = \frac{H_{\infty}^3 \Omega_M}{s_0^3 \Omega_{\Lambda}}, \qquad
c_2 = \frac{H_{\infty}^4 \Omega_R}{s_0^4 \Omega_{\Lambda}}, \qquad
c_3 = \frac{H_{\infty}^2 \Omega_K}{s_0^2 \Omega_{\Lambda}},
\end{equation}
and a recurring denominator term:
\begin{equation}\label{eq:coeff_Kc}
\mathcal{K}_c = k^2 + 3c_3.
\end{equation}

For brevity in presenting $\mathbf{X}_1$, we define the following intermediate expressions:
\begin{align}\label{eq:ABCD_defs}
    A &= \frac{6k^6+26k^4c_3+288c_2c_3+12k^2(2c_3^2-7c_2)}{45\mathcal{K}_c}, \\
    B &= \frac{c_1}{\mathcal{K}_c} \left( -\frac{9}{2}\Delta\eta + \frac{2k^2-3c_3}{4}(\Delta\eta)^3 \right), \\
    C &= \frac{c_1}{\mathcal{K}_c} \left( -\frac{27}{2}\Delta\eta - \frac{3(k^2+12c_3)}{4}(\Delta\eta)^3 \right), \\
    D &= \frac{3k^2+4c_3}{120}.
\end{align}
The matrix $\mathbf{X}_1$ is then given by:
\begin{widetext}
\setlength{\arraycolsep}{1.5pt} 
\begin{equation}\label{eq:X1_matrix_full}
\mathbf{X}_1=
\begin{pmatrix}
0 &
\frac{c_1}{\mathcal{K}_c} \left( -\frac{3}{2}\Delta\eta - \frac{c_3}{4}(\Delta\eta)^3 \right) &
\frac{c_2}{\mathcal{K}_c} \left( 6\Delta\eta + \frac{k^2+8c_3}{5}(\Delta\eta)^3 \right) &
\frac{c_1}{\mathcal{K}_c} \left( -\frac{9}{2}\Delta\eta - \frac{3(k^2+4c_3)}{4}(\Delta\eta)^3 \right) \\
0 &
\frac{c_1}{\mathcal{K}_c} \left( -\frac{3}{2}\Delta\eta - \frac{c_3}{4}(\Delta\eta)^3 \right) &
\frac{c_2}{\mathcal{K}_c} \left( 6\Delta\eta + \frac{k^2+8c_3}{5}(\Delta\eta)^3 \right) &
\frac{c_1}{\mathcal{K}_c} \left( -\frac{9}{2}\Delta\eta - \frac{3(k^2+4c_3)}{4}(\Delta\eta)^3 \right) \\
1 - \frac{1}{6}k^2 (\Delta\eta)^2 &
\frac{c_1}{\mathcal{K}_c} \left( -6\Delta\eta - \frac{3c_3-2k^2}{3}(\Delta\eta)^3 \right) &
-\frac{4k^4+12k^2c_3-72c_2}{3\mathcal{K}_c}\Delta\eta + A(\Delta\eta)^3 &
\frac{c_1}{\mathcal{K}_c} \left( -18\Delta\eta - (k^2+12c_3)(\Delta\eta)^3 \right) \\
0 &
1 + B &
\frac{c_2}{\mathcal{K}_c} \left( 18\Delta\eta + \frac{24c_3-7k^2}{5}(\Delta\eta)^3 \right) &
\frac{1}{2}k^2(\Delta\eta)^2 + C \\
\frac{1}{4}\Delta\eta - D(\Delta\eta)^3 &
-\frac{3c_1}{2\mathcal{K}_c}(\Delta\eta)^2 &
1 - \frac{3k^4+13k^2c_3+12(c_3^2-5c_2)}{10\mathcal{K}_c}(\Delta\eta)^2 &
-\frac{9c_1}{2\mathcal{K}_c}(\Delta\eta)^2 \\
0 &
-\frac{3c_1}{2\mathcal{K}_c}(\Delta\eta)^2 &
\frac{6c_2}{\mathcal{K}_c}(\Delta\eta)^2 &
-\Delta\eta - \frac{9c_1}{2\mathcal{K}_c}(\Delta\eta)^2 - \frac{c_3}{6}(\Delta\eta)^3
\end{pmatrix};
\end{equation}
\setlength{\arraycolsep}{5pt} 
\end{widetext}
The matrix $\mathbf{X}_2$ is given by:
\begin{widetext}
\setlength{\arraycolsep}{1.5pt} 
\begin{equation}\label{eq:X2_matrix_full}
\mathbf{X}_2 =
\begin{pmatrix}
\frac{1}{15}k^2(\Delta\eta)^2 - \frac{k^2(15k^2+14c_3)}{3150}(\Delta\eta)^4 &
-\frac{4k^2c_1}{15\mathcal{K}_c}(\Delta\eta)^3 &
\frac{8}{15}k^2\Delta\eta - \frac{4k^2(30k^4+118k^2c_3+84(c_3^2-5c_2))}{1575\mathcal{K}_c}(\Delta\eta)^3 &
-\frac{4c_1k^2}{5\mathcal{K}_c}(\Delta\eta)^3 \\
-\frac{1}{105}k^3(\Delta\eta)^3 &
\frac{k^3c_1}{35\mathcal{K}_c}(\Delta\eta)^4 &
-\frac{4}{35}k^3(\Delta\eta)^2 + \frac{k^3(30k^4+118k^2c_3+84(c_3^2-5c_2))}{3675\mathcal{K}_c}(\Delta\eta)^4 &
\frac{3k^3c_1}{35\mathcal{K}_c}(\Delta\eta)^4
\end{pmatrix}.
\end{equation}
\setlength{\arraycolsep}{5pt} 
\end{widetext}

\let\oldbibitem\bibitem
\renewcommand{\bibitem}{%
    \renewcommand{\doi}[1]{doi: ##1}
    \let\bibitem\oldbibitem
    \oldbibitem
}

\bibliography{Reference}

\end{document}